# The Memory Controller Wall: Benchmarking the Intel FPGA SDK for OpenCL Memory Interface

Hamid Reza Zohouri*†[1], Satoshi Matsuoka*‡
*Tokyo Institute of Technology, †Edgecortix Inc. Japan, ‡RIKEN Center for Computational Science (R-CCS)
{zohour.h.aa@m, matsu@is}.titech.ac.jp

***Abstract*—Supported by their high power efficiency and recent advancements in High Level Synthesis (HLS), FPGAs are quickly finding their way into HPC and cloud systems. Large amounts of work have been done so far on loop and area optimizations for different applications on FPGAs using HLS. However, a comprehensive analysis of the behavior and efficiency of the memory controller of FPGAs is missing in literature, which becomes even more crucial when the limited memory bandwidth of modern FPGAs compared to their GPU counterparts is taken into account. In this work, we will analyze the memory interface generated by Intel FPGA SDK for OpenCL with different configurations for input/output arrays, vector size, interleaving, kernel programming model, on-chip channels, operating frequency, padding, and multiple types of overlapped blocking. Our results point to multiple shortcomings in the memory controller of Intel FPGAs, especially with respect to memory access alignment, that can hinder the programmer's ability in maximizing memory performance in their design. For some of these cases, we will provide work-arounds to improve memory bandwidth efficiency; however, a general solution will require major changes in the memory controller itself.**

## I. Introduction

Due to the ever-increasing need for more power efficient means of accelerating computational workloads, FPGAs are now being considered as a potentially more-efficient alternative to GPUs [1]. With the programmability challenge of using FPGAs being significantly alleviated with adoption of modern C/C++ and OpenCL-based High-Level Synthesis (HLS) techniques [2, 3] in the past few years, research groups, large IT companies and even cloud providers are more rapidly adopting these devices to accelerate different workloads. In line with this trend, a lot of effort has been made to optimize different applications on FPGAs. However, most such efforts have been focused on kernel-level optimizations such as optimizing loop initiation interval, reducing area usage, and maximizing parallelism, while the critical effect of external memory and its controller on performance is largely overlooked. It has been shown that due to the low external memory bandwidth and low byte to FLOP ratio of modern FPGAs, it is frequently not possible to fully utilize the compute performance of these devices for typical applications [1, 4]. Moreover, even this limited external memory bandwidth cannot always be efficiently utilized due to limitations in the memory controller [4, 5, 6]. In this work, we will focus on analyzing the efficiency of the external memory controller on Intel FPGAs for OpenCL-based designs under different configurations. Our contributions are as follows:

- We create an OpenCL-based memory benchmark suite for Intel FPGAs with support for different memory access patterns and multiple compile-time and run-time-configurable parameters.
- Using our benchmark suite, we perform a comprehensive analysis of the memory controller performance and efficiency on Intel FPGAs with different configurations for input/output arrays, vector size, interleaving, kernel programming model, on-chip channels, operating frequency, padding, and multiple types of blocking.
- We outline one performance bug in Intel's compiler, and multiple deficiencies in the memory controller, leading to significant loss of memory performance for typical applications. In some of these cases, we provide work-arounds to improve the memory performance.

[1] This work was conducted at Tokyo Institute of Technology

## II. Methodology

### A. Memory Benchmark Suite

For our evaluation, we develop an open-source benchmark suite called FPGAMemBench, available at https://github.com/zohourih/FPGAMemBench. The benchmark suite covers multiple memory access patterns that are regularly found in HPC applications with different number of input and output arrays. Moreover, multiple compile-time and run-time-configurable parameters are provided that can affect the memory access pattern and/or performance.

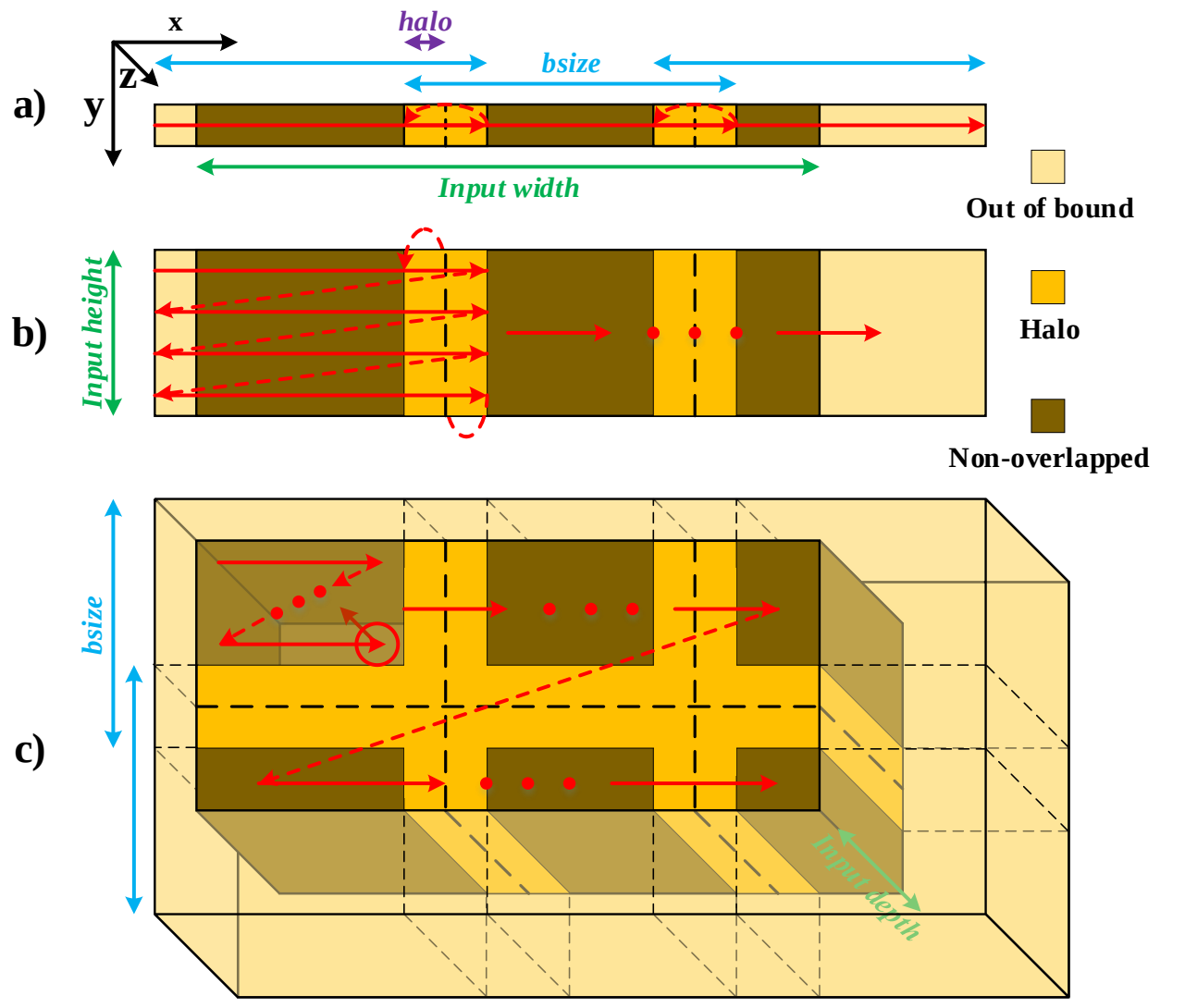

Fig. 1. Access pattern for a) 1D, b) 1.5D, and c) 2.5D overlapped blocking

Three main classes of memory access patterns are supported by our benchmark suite. The first class, shown in Fig. 1 a), implements 1D overlapped blocking. In this case, 1D input and output arrays are used and, as shown by the red arrows, memory accesses happen in a block-based manner starting from the address $-halo$, continuing linearly in a vectorized fashion to $bsize - halo$, and then the next block starts from address $csize = bsize - 2 \times halo$. Here, $bsize$ is the block size which is a compile-time variable, and the blocks are overlapped by $2 \times halo$ indexes, with $halo$ being a run-time variable that controls halo/overlap size. A halo size of zero results in simple sequential reading/writing with no overlapping. The second and third classes implement 1.5D and 2.5D overlapped spatial blocking, respectively, that are widely used in 2D and 3D stencil computation [5, 6, 7, 8, 9, 10]. For the 1.5D class, the $x$ dimension is blocked and memory accesses are streamed row by row until the last index in the $y$ dimension, before moving to the next block (Fig. 1 b)). Similarly, in the 2.5D class, the $x$ and $y$ dimensions are blocked, while the accesses are streamed plane by plane in the $z$ dimension. In the 1.5D class, blocks are overlapped by $2 \times halo$ columns, while in the 2.5D class, they are overlapped by $2 \times halo$ rows and columns. All out-of-bound accesses (negative row, column and indexes, and out-of-bound indexes in the rightmost and bottommost blocks if input dimensions are not divisible by *csize*) are avoided in the kernels using conditional statements.

Each class of memory access patterns has both a *standard* and a *channelized* implementation. The former performs all memory operations in the same kernel, while the latter uses one kernel for reading and one kernel for writing, with data being passed between them using on-chip channels. For the 1D blocking class, each kernel file includes both an NDRange (NDR) and a Single Work-item (SWI) implementation of 5 kernels with different number of input and output arrays. As shown in Fig. 2, four global memory buffers are used in total and the supported array configurations are R1W0, R1W1, R2W1, R3W1, and R2W2; here, R and W refer to "read" and "write" respectively, and the numbers following them show the number of arrays that are read or written. To prevent the R1W0 kernel from being optimized out by the compiler, a one-time memory write is also added in the kernel. The R1W1 and R2W2 kernels implement one and two direct buffer copies, respectively, while the R2W1 and R3W1 kernels sum up the input values to generate the output. The 1.5D and 2.5D blocking classes support all the above array configurations except R1W0. All Single Work-item kernels use collapsed loops with the exit condition optimization from [4, 5, 6] for best timing and hence, are constructed as a doubly-nested loops, with the fully-unrolled innermost loop having a trip count equal to the vector size, and the outer loop having an initiation interval of one. In the NDRange kernels, the work-groups have the same number of dimensions as the input, and memory access coalescing is performed using loop unrolling. Even though the same behavior could be achieved in this kernel model using the compiler's built-in SIMD attribute with no performance difference, it was avoided since it would have resulted in multiple channel call sites in the channelized kernels. The "float" data type is used in all our kernels.

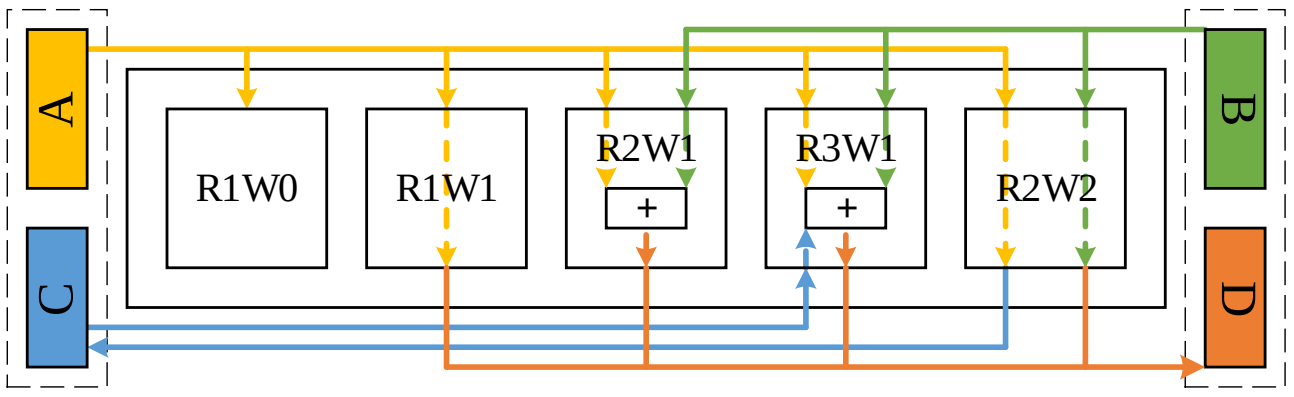

Fig. 2. Overview of kernels and arrays

To allow testing different alignment configurations, our benchmark suite supports array padding, where the start of all global arrays is padded by a run-time-configurable number of indexes (*pad*), moving the starting address to $pad - halo$. Moreover, row-wise padding is supported for the 1.5D and 2.5D classes, and column-wise padding is supported for the 2.5D class. As final note, Intel FPGA SDK for OpenCL provides no user control over the final operating frequency of the design and automatically tries to maximize the operating frequency. This limitation creates a significant hurdle for benchmarking since it requires that the effect of the variable operating frequency, which is not necessarily linear, to be also taken into account. To overcome this issue, we created a work-around to override the final operating frequency of the design to a user-defined value, the details of which are provided in our open-source benchmark repository. However, this work-around is usable only if the design's maximum-achievable operating frequency is above the user-defined value.

### B. *Hardware and Software Setup*

For our evaluation, we use a Nallatech 385A board equipped with an Arria 10 GX 1150 FPGA and two banks of DDR4 memory running at 2133 MT/s. Each memory bank is connected to the FPGA via a 64-bit bus, providing a theoretical peak memory bandwidth of 34.128 GB/s[1]. The memory controller of the FPGA runs at $1/8$ of the memory frequency (266.666 MHz). We also use an old NVIDIA Tesla K20X and a modern Tesla V100 SXM2 for comparison. The peak memory bandwidth of these GPUs is 249.6 and 897.0 GB/s, respectively. All our machines use an updated version of CentOS 7, and we use GCC v7.4.0 for compiling host codes and CUDA v10.0 for GPUs. The latest BSP released by Nallatech for our board is only compatible with Quartus v17.1.0 and hence, we cannot use the latest version of Intel's compiler for our evaluation. However, since support for BSP backward compatibility with up to 2 lower major versions of Quartus was added in Intel FPGA SDK for OpenCL v18.1, it is possible to use the v17.1.0 BSP with all 18.1.x versions of the compiler. Hence, we use Intel FPGA SDK for OpenCL v18.1.2, Quartus v17.1.2, and Nallatech BSP v17.1.0 in our evaluation. Even though we encountered multiple cases where our kernels compiled using this setup hung during execution for no clear reason, we continued using this version so that our evaluation reflects the latest-possible version of the compiler, and worked around the hangs by changing the placement and routing seed. Since the issue is likely caused by the backward compatibility feature, we would *not* recommend this setup for a production environment. We employed test functions in every case to ensure the correctness of the kernel outputs.

### C. *Benchmark Settings*

For our FPGA evaluation, we use array sizes of 1 GiB, for a total memory usage of 4 GiB. All input arrays are initialized with random single-precision floating-point numbers. For the 1D and 1.5D kernels, *bsize* is set to 1024, and for 2.5D, it is set to 256×256. Moreover, in the NDRange kernels, the work-group size is set to *bsize* divided by vector size. None of these parameters have a direct effect on performance. All our kernels are compiled with the manually-defined operating frequency of 266.666 MHz (same as the memory controller). If a kernel does not meet the timing requirement for this frequency, we change the compilation seed and/or increase the target operating frequency in the OpenCL compiler until the requirement is met. Every kernel is run 5 times and average run time is measured. For timing measurement, we only measure kernel run time (excluding PCI-E transfer) using the high-precision clock_gettime() function.

For GPU evaluation, we use the NDRange kernels of our benchmark suite with a vector size of 1 (vectorization by loop unrolling is not applicable to GPUs), and adjust the work-group size (equal to *bsize* here) for each GPU to maximize its performance. It should be noted that in practical applications, unless the number of registers per thread is a limitation, the work-group size on GPUs can be changed freely to maximize memory performance – something that does *not* apply to the vector size on FPGAs due to area overhead.

[1] GB = $10^9$ B, GiB = $2^{30}$ B

## III. Evaluation

### A. No Overlapping

As the first step, we use the 1D class with blocking disabled ($halo = 0$) to evaluate standard sequential accessing and determine the effect of basic parameters.

*1) Effect of Vector Size:* To evaluate the effect of vector size, we use the NDRange variation of our kernels with the default compiler setting that interleaves the global memory arrays between the external memory banks. Since the total size of the memory bus is 128 bits (2 × 64 bits) and both the kernel and memory controller run at $1/8$ of the external memory frequency, a total access size of 1024 bits (8 × 128 bits = 32 floats) per clock should be enough to saturate the memory bandwidth. Fig. 3 shows the performance of all array configurations with all power of two vector sizes from 1 to 32. Non-power-of-two vector sizes were avoided since the size of memory ports generated by the compiler are always increased to the nearest bigger power of two and extra bytes are masked out, resulting in significant waste of memory bandwidth.

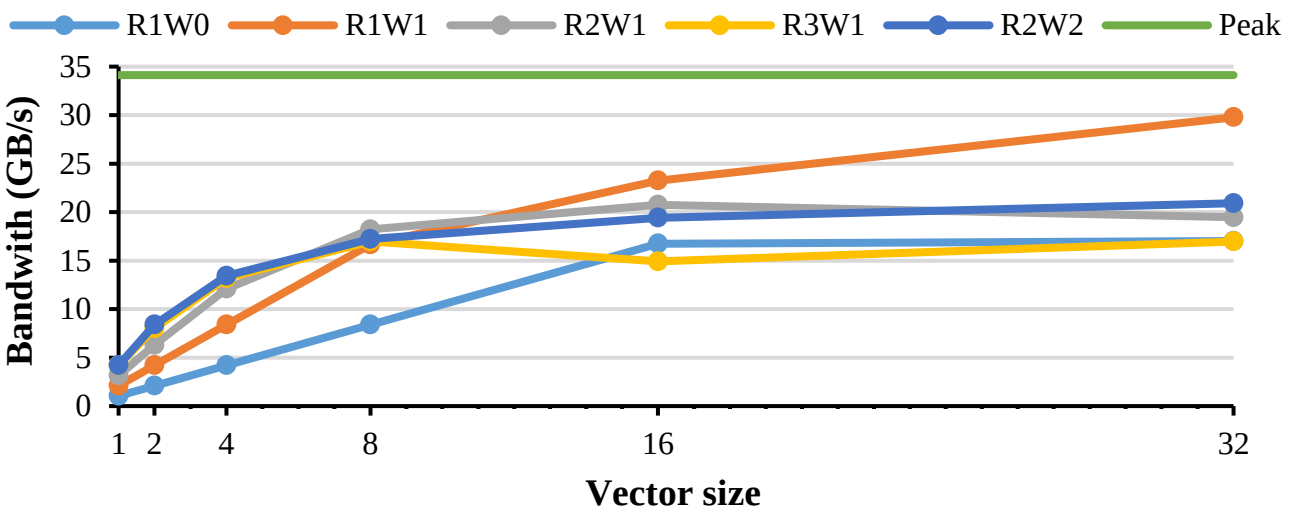

Fig. 3. Performance scaling with vector size with interleaving enabled

For the R1W0 kernel, we expect performance to scale linearly with vector size and saturate the memory bandwidth at a vector size of 32. Based on the results, the performance scales linearly up to a vector size of 16, but does not scale beyond that. This shows that **even with interleaving, memory performance will not scale to more memory banks than there are global arrays in the kernel, regardless of vector size.** For the R1W1 case, we expect memory performance to increase linearly until it saturates at a vector size of 16. However, in practice, performance scaling becomes non-linear above the vector size of 8; even though performance keeps increasing up to a vector size of 32, the memory bandwidth is still not saturated. For other array configurations, performance scales linearly only up to a vector size of 4, and stops scaling at a vector size of 16, while the maximum achieved memory bandwidth is close to only half the peak bandwidth. Based on our results, **the memory controller/interface performs poorly when more arrays are used than there are memory banks and vector size is large**. Our performance results for small vector sizes also show that **no run-time coalescing is performed by the OpenCL memory interface on Intel FPGAs.**

Fig. 4 shows the efficiency of memory controller for the aforementioned benchmarks. The efficiency in this case is the ratio of measured to expected throughput; the latter is calculated by multiplying the number of arrays by vector size by the operating frequency (266.666 MHz) by the data type size (4 bytes). In case the expected throughput is higher than the peak throughput, we use the peak throughput. Based on the results, the memory controller achieves ~100% efficiency with a vector size of 1 for all array configurations, while ~100% efficiency is achieved with a vector size of 2 for all configurations except R3W1. For larger vector sizes up to 8, only the simpler R1W0 and R1W1 configurations achieve full efficiency, but the controller efficiency starts falling drastically after that. Efficiency still increases with a vector size of 32 for some of the configurations since the expected throughput is capped by the peak theoretical bandwidth in these cases, while the actual increase in measured bandwidth is minimal (Fig. 3). Moreover, the area/power efficiency would decline in such cases since the increase in memory performance will be smaller than the increase in area usage.

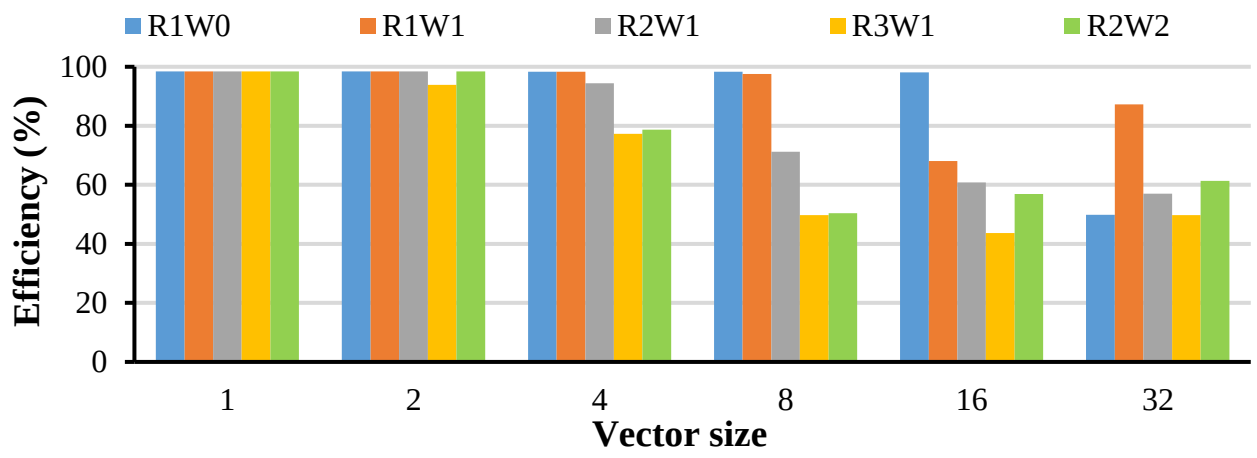

Fig. 4. Memory controller efficiency with interleaving enabled

*2) Effect of Interleaving:* Fig. 5 shows the performance of all array configurations with interleaving disabled. In this case, we experimentally found that highest performance is achieved for every case if the four arrays are banked as shown by the hachured boxes in Fig. 2. With interleaving disabled, in nearly every case, performance is either the same or higher compared to the same configuration with interleaving enabled. Here, the R1W1 configuration achieves near-linear performance scaling up to a vector size of 16 and nearly saturates the memory bandwidth. Since performance does not improve beyond this point, **we believe the ~32.6 GB/s we measure for this configuration is the peak achievable memory bandwidth on this board, which puts the maximum-achievable memory bandwidth efficiency at 95.7%.** Moreover, **our results do *not* show any tangible performance benefit for using the default compiler flow with interleaving enabled, except allowing allocation of buffers that are larger than the size of one memory bank on the FPGA board.**

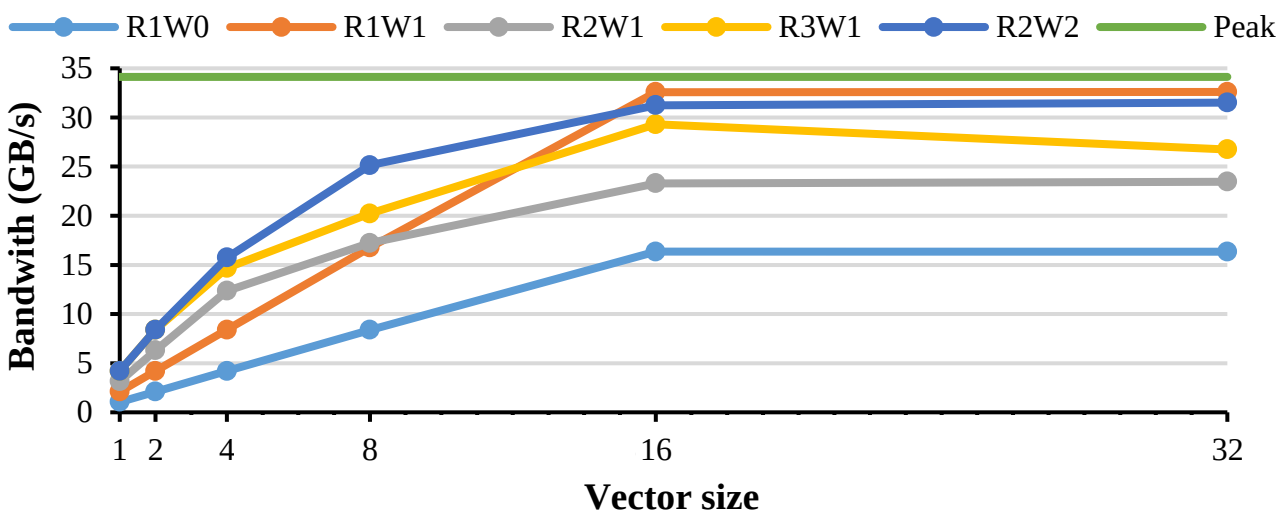

Fig. 5. Performance scaling with vector size with interleaving disabled

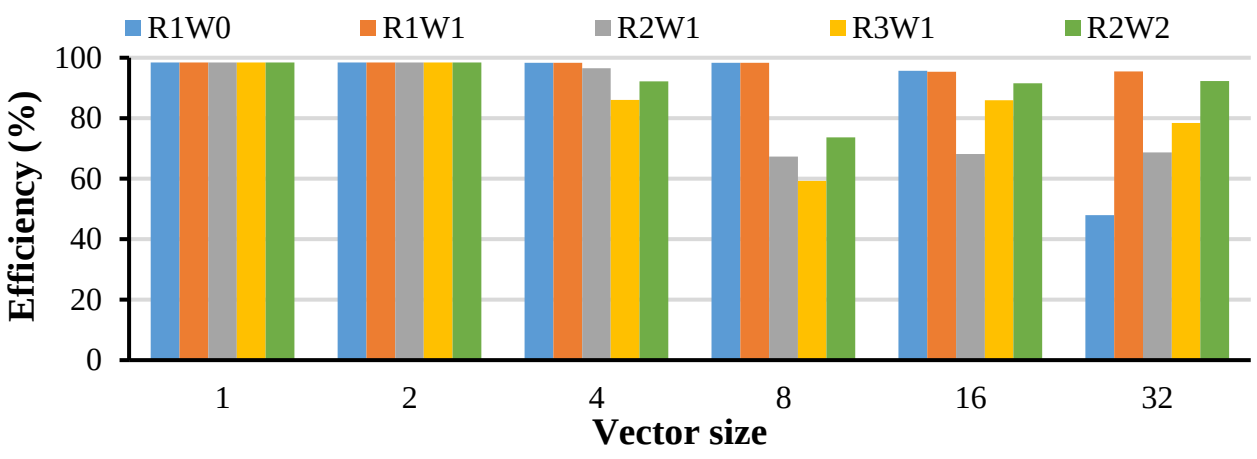

Fig. 6. Memory controller efficiency with interleaving disabled

Fig. 6 shows the memory controller efficiency with interleaving disabled. The expectation is that for the double and triple-array kernels, a vector size of 16, and for the quadruple-array ones, a vector size of 8 should be enough to saturate the memory bandwidth. However, in practice, this is only achieved for the double-array case. For the R2W2 case, double the necessary vector size can achieve reasonable memory throughput (at the cost of lower area and power efficiency); however, for the remaining multi-array cases, even a vector size of 32 does not allow achieving more than 85% of the peak memory bandwidth, with the R2W1 case being the worst at ~70% bandwidth efficiency.

Our results show that very specific requirements need to be met to achieve the highest possible memory throughput on Intel FPGAs programmed using OpenCL. Specifically, **the number of global arrays in the design must be same the as the number of memory banks, vector size should be 16, and interleaving should be disabled and arrays manually partitioned between the memory banks**. Reasonable memory throughput can still be achieved if the number of arrays is more, but divisible by the number of banks. However, such cases will require larger vector sizes than should be necessary to achieve such performance, leading to lowered area and power efficiency. It is clear that following such strict patterns to achieve high bandwidth efficiency is not always feasible, which means the efficiency of the memory controller will become a major bottleneck in a large class of OpenCL-based designs on Intel FPGAs that cannot satisfy these requirements. As a work-around, if the number of arrays is more than the number of memory banks but the size and access pattern of the arrays is exactly the same, it is possible to merge the all the read arrays into one *array of structs* structure, and all the write arrays to another, resulting in an R1W1 configuration. Here, the same index of all the arrays can be read/written using one memory access, and the memory accesses can be further vectorized to achieve maximum memory bandwidth efficiency. However, this work-around will require that the number of read and write arrays be a power of two or else the total size of the memory ports will not be a power of two after the arrays are merged, leading to waste of memory bandwidth due to bit masking. For other configurations (e.g. R3W1), no straightforward work-around can be considered except improving the memory controller itself.

As a comparison, Fig. 7 shows the memory controller efficiency (measured performance relative to peak) of our kernels on NVIDIA Tesla K20X and V100 GPUs. We experimentally find work-group sizes of 128 and 512 to achieve highest performance on these GPUs, respectively. On the old K20X GPU, with the exception of the single-array R1W0 kernel, the other kernels achieve similar performance with a memory bandwidth efficiency of ~75%. The low performance of the R1W0 kernel in this case is likely due to thread divergence caused by the branching over the write operation that is added to prevent the kernel from being optimized out, rather than due to the memory controller efficiency. On the modern V100 GPU, however, reasonable memory bandwidth efficiency can be achieved for all kernels, with the minimum efficiency being 88%. **These results show that memory performance on NVIDIA GPUs has much less variation compared to Intel FPGAs with different array configurations, pointing to a noticeably more efficient memory controller on these devices.**

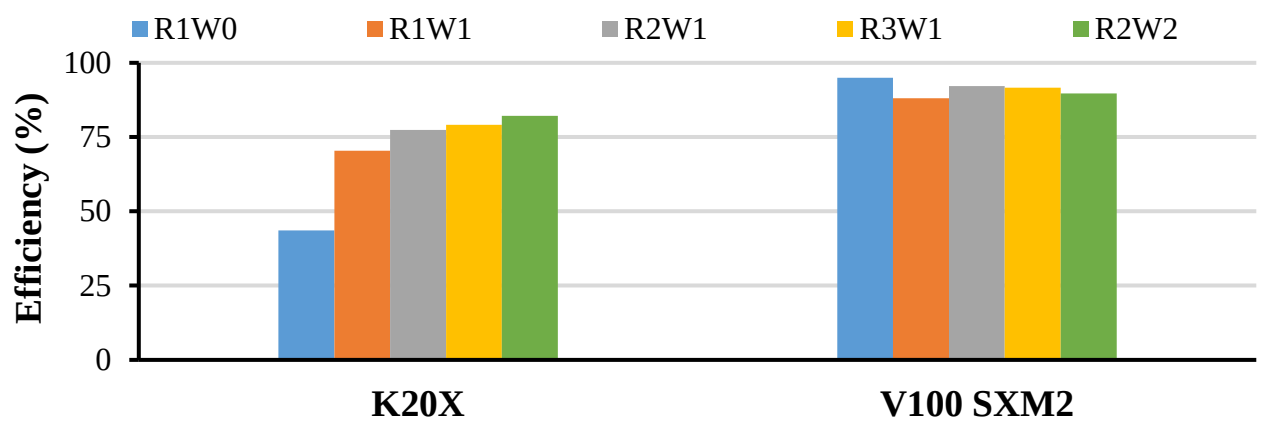

Fig. 7. Memory contoller efficiency on GPUs

*3) Effect of Kernel Programming Model:* Fig. 8 shows the performance of the Single Work-item version of our kernels with interleaving disabled. Compared to the results of the NDRange kernels (Fig. 5), no visible difference in memory performance can be observed. We did not observe any noticeable difference with interleaving enabled, either. **Hence, we conclude that with respect to memory performance, neither of the kernel programming models are preferred.**

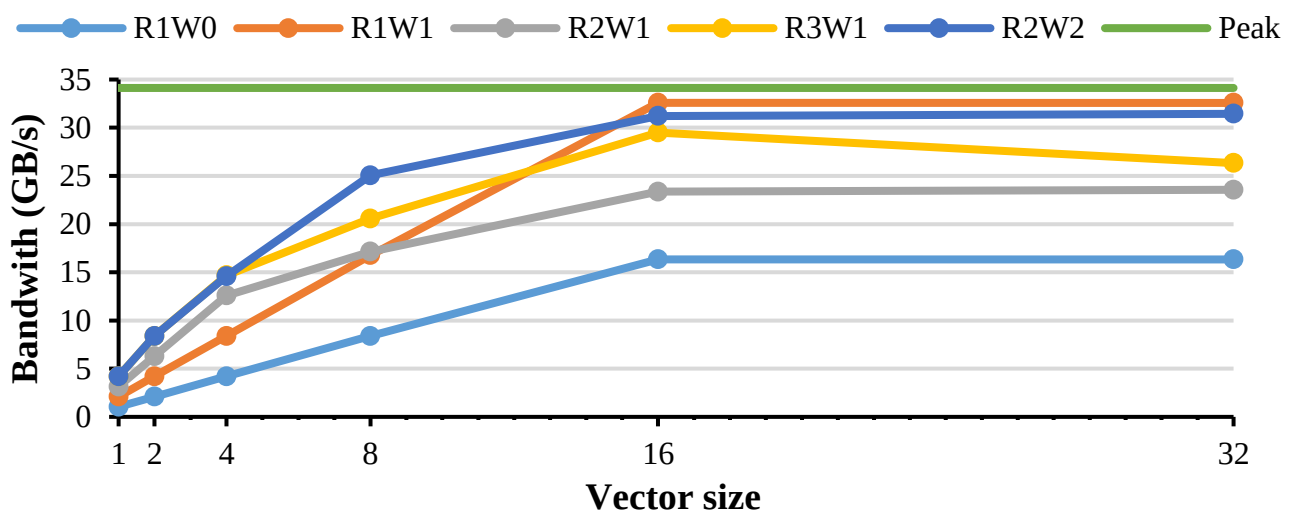

Fig. 8. Non-intereleaved performance scaling with SWI kernels

*4) Effect of Channels:* Fig. 9 shows the performance of the channelized variation of our kernels with interleaving disabled for both kernel programming models. The channel depth in this case is set to 16 which is reasonable for the regular kernels studied here and is also small enough to avoid using Block RAMs. Our expectation is that the existence of channels in the kernels should not have any effect on memory performance since it merely increases the pipeline latency. However, our results paint a different picture. Comparing Fig. 9 top and Fig. 5 shows that in multiple cases, there are minor performance improvements in the channelized NDRange kernel, especially for the triple and quadruple-array kernels where the amount of contention on the memory bus is higher. Similar minor improvement can be seen also with interleaving disabled (results are omitted for brevity). We speculate that the existence of the channels is increasing the pipeline's ability in absorbing stalls from external memory reads and preventing them from propagating to the memory write operation, leading to minor performance improvements. However, the case of the Single Work-item kernel is opposite: the performance of the channelized kernels is lower in most cases, with the difference being as large as 20% for the R1W1 case (Fig. 9 bottom vs. Fig. 8). Similar performance regression is also seen with interleaving disabled. Increasing the channel depth up to an unreasonable value of 512 had no effect on performance, eliminating this parameter as the possible cause of the problem. Using older versions of the compiler, both v17.1.0 and v16.1.2 exhibited the same problem, but the performance loss was largest with 17.1.0 and smallest with 16.1.2. We also reproduced this performance regression on our Stratix V board using different versions of the compiler. **Our results point to a possible flaw in the compiler which affects a large range of applications, from simple single-input single-output applications that use on-chip channels to simplify kernel design, to highly-optimized designs relying on replicated autorun kernels to implement rings and systolic arrays of processing elements.** In fact, we encountered this exact issue in our previous work [4, 5, 6], but we could not find the cause and hence, avoided using the v17.0+ versions of the compiler due to lower performance. We have reported this issue to Intel, and they have been able to reproduce it. However, a fix, if any, will likely never be available on boards with Arria 10 and older FPGAs due to lack of BSP updates. As a work-around, for designs that separate memory access kernels from compute (similar to [5, 6]), the memory kernels can be converted to NDRange to avoid this problem.

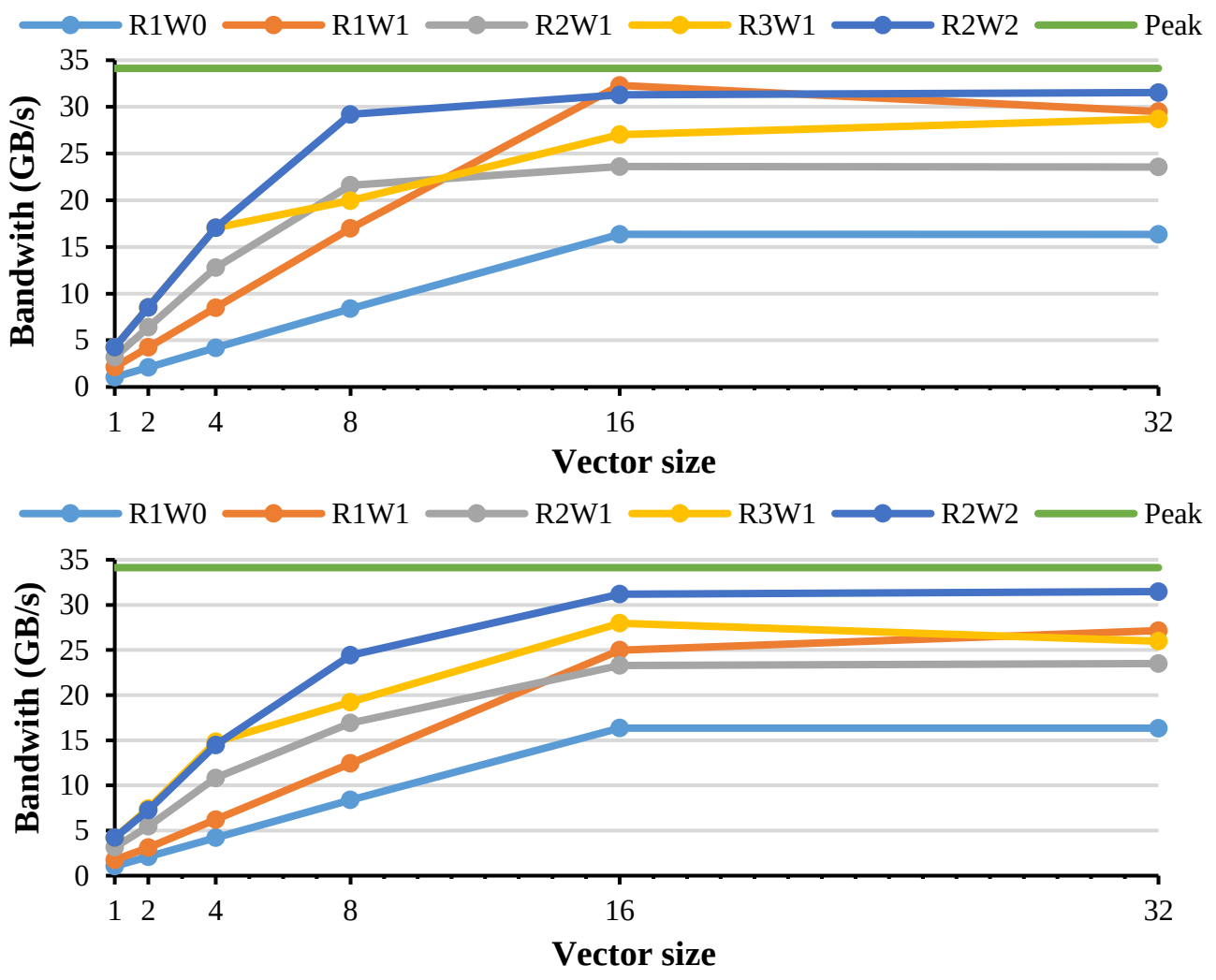

Fig. 9. Channelized kernel performance; top: NDR, bottom: SWI

*5) Effect of Operating Frequency:* Fig. 10 shows the performance of our evaluated kernels with multiple different operating frequencies, including one above the memory controller frequency (300 MHz), and interleaving disabled; the red line marks the memory controller frequency. We chose a vector size of 8 for this experiment to prevent the effect of operating frequency from being hidden by saturation of the memory bandwidth. Based on the results, **performance of all the kernels scales linearly with operating frequency, until a specific threshold is reached above which performance starts *decreasing* with higher frequency due to increased contention on the memory bus.** For the single and double-array kernels, this threshold is not reached up to 300 MHz since the achieved memory bandwidth for these kernels with a vector size of 8 is still far from the peak memory bandwidth. For the rest of the configurations, however, performance saturates between 200 to 250 MHz and decreases after that. Similar effects were observed with interleaving enabled or other vector sizes using either of the two kernel programming models. As such, seed and $F_{max}$ sweeping can be used to *increase* operating frequency, and consequently, memory performance, with small resource overhead [4, 5, 6]. On the other hand, for multi-array kernels with wide vectors where higher operating frequency can decrease performance, such as the case discussed in Section 4.3.1.6 of [4], our work-around to override operating frequency can be used to increase memory performance by *reducing* operating frequency instead.

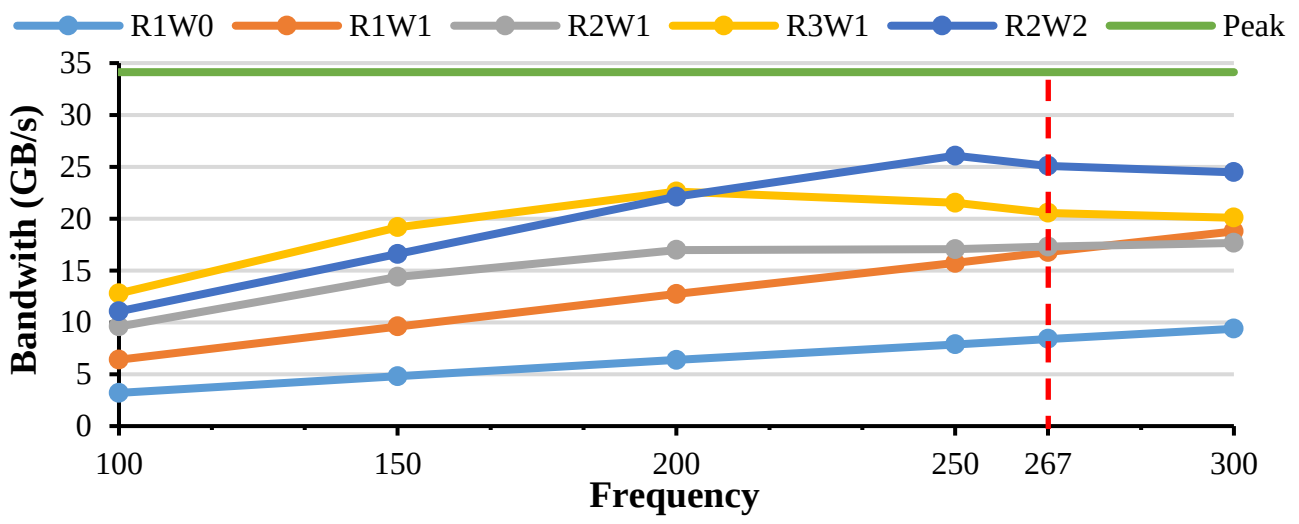

Fig. 10. Performance scaling with operating frequency for vector size of 8

*6) Effect of Alignment:* To determine the effect of memory access alignment, we utilize basic array padding which adds a fixed offset to the memory addresses. Fig. 11 shows the performance of the NDRange R1W1 kernel with interleaving disabled and padding sizes ranging from 0 (no padding) to 32 floats. Each line in this figure represents a specific vector length.

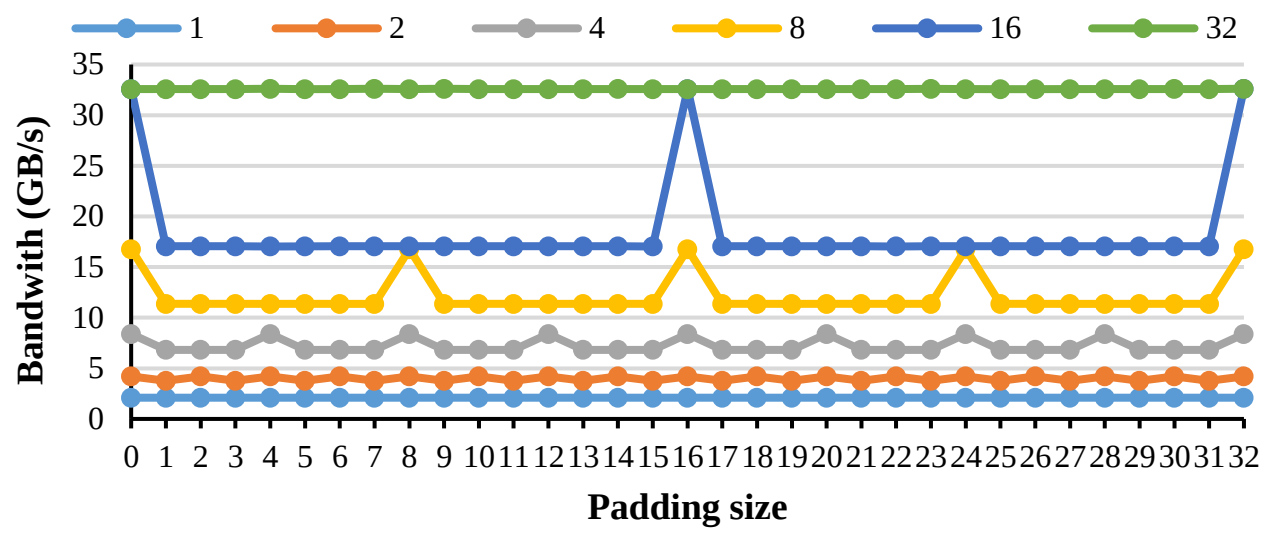

Fig. 11. R1W1 performance with padding for different vector sizes

**Our results show that the memory controller on the Intel Arria 10 FPGA is not capable of performing any memory access realignment at all, losing up to ~50% of its performance due to unaligned memory accesses.** We observed the same patterns regardless of kernel programming model, compiler version, interleaving state, and array configuration. The only difference in these cases is the amount of performance that is lost; e.g. if the memory controller is oversubscribed using larger vector sizes than are necessary to saturate the memory bandwidth, the amount of performance lost due to memory access misalignment gets smaller. We also observed the same behavior on our Stratix V board, and expect that the same will also apply to Stratix 10 GX FPGAs with DDR memory. We previously suggested that a memory address alignment of 512 bits is required to achieve maximum memory bandwidth on Intel FPGAs [5, 6]. However, our results here show that the required alignment is not fixed and is equal to the vector size, making the alignment requirement stricter as the vector size increases. With all these, vector size of 32 (and possibly higher values) does not seem to be affected by memory access alignment; albeit, as discussed before, using such large vector sizes is not area efficient and might not even be feasible in many cases due to limited FPGA resources.

In comparison, [11] reports that even though the same behavior was also exhibited by early Nvidia Tesla cards released over 10 years ago, it was largely addressed with the release of Tesla C2050. The effect of padding on our GPUs is shown in Fig. 12. Performance variations relative to padding can be observed on both GPUs, with maximum performance achieved at padding sizes that are a multiple of 8, pointing to a required alignment size of 256 bits. Here, the Tesla V100 GPU which uses HBM memory exhibits higher performance variation than K20X which uses GDDR memory, but the maximum loss of performance with misalignment on this GPU is only ~13% – still far lower than our FPGA.

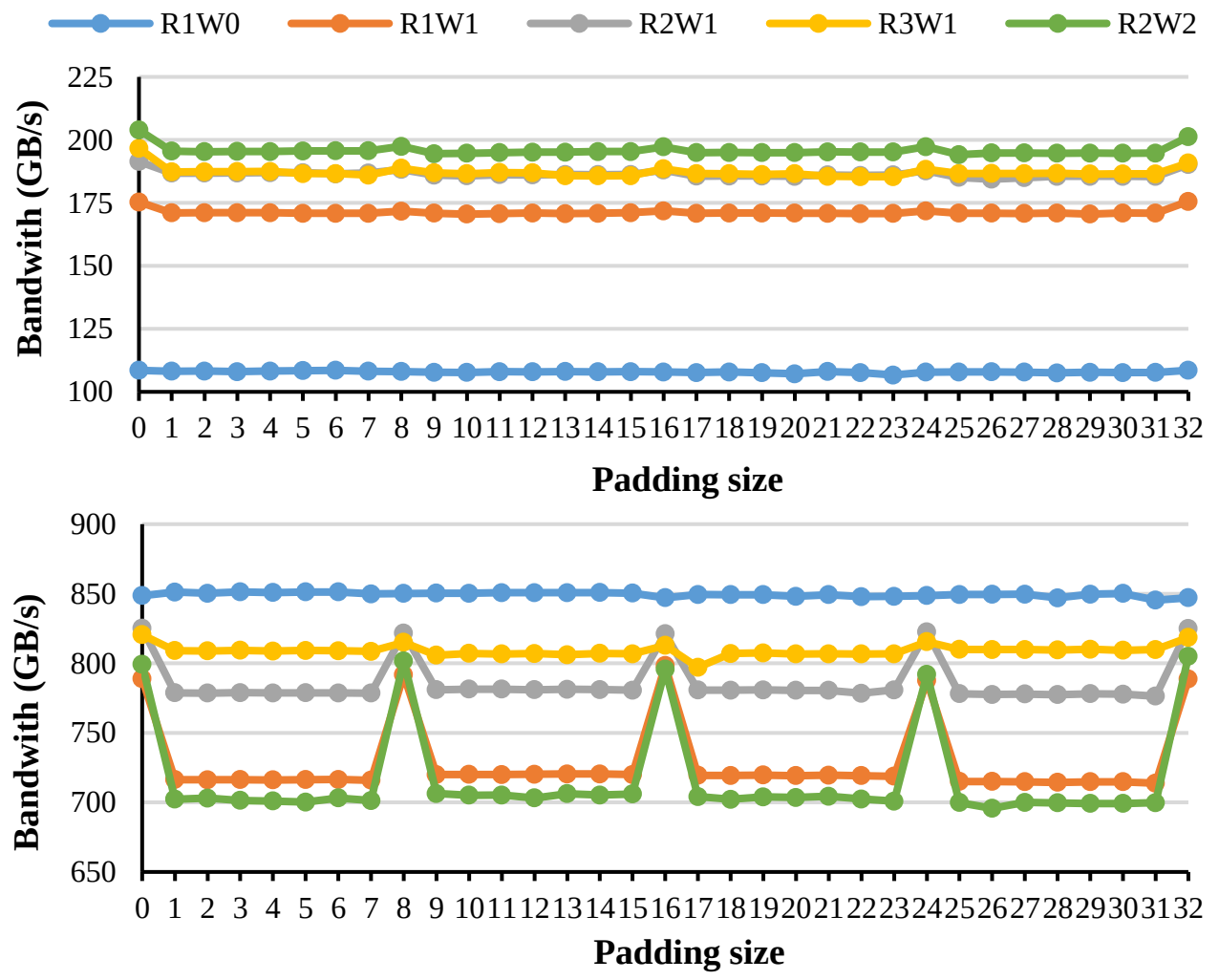

Fig. 12. Effect of padding on GPUs; top: Tesla K20X, bottom: Tesla V100

## B. 1D Blocking

In this section, we enable overlapping and analyze its effect alongside with other parameters.

*1) Effect of Overlapping:* Fig. 13 shows the effect of halo size (different bar color) on the R1W1 kernel for different vector sizes (different clusters) and interleaving disabled. The vector size of 1 is skipped here since performance in that case is not affected by alignment. Moreover, we limited the halo sizes to powers of two since alignment characteristics for odd halo sizes is the same as the halo size of 1, and for even values, performance only changes depending on the biggest power of two that the halo size is divisible by. E.g. performance for halo sizes of 6 and 14 is the same as halo size of 2, 12 and 20 is the same as 4, and so on. The array size for this experiment has been adjusted for every halo size to the nearest size to 1 GiB where the number of array indexes is a multiple of *csize*, to avoid skewing bandwidth measurement by out-of-bound processing in the last blocks (Fig. 1). Redundant memory accesses are also counted in bandwidth calculation here.

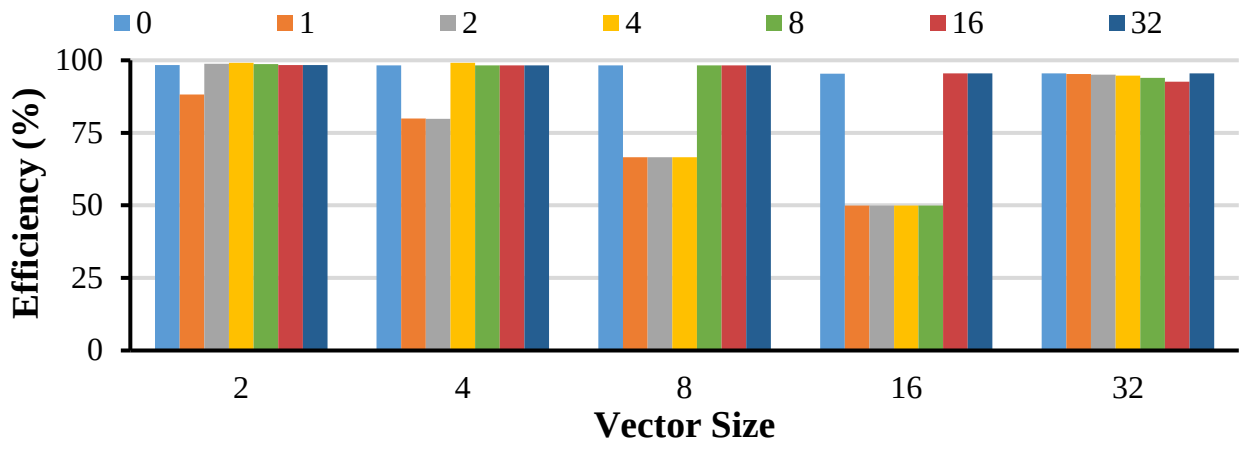

Fig. 13. Memory controller efficiency with overlapping for R1W1 kernel

Based on the results, high performance efficiency can be achieved as long as the halo size is a multiple of vector size, but if this condition is not met, performance can suffer by up to 50%. Here, if the halo size is not a multiple of vector size, the starting address of the first block and every subsequent access inside the block will not be aligned (Fig. 1 a)). Moreover, since the block size in our case is a power of two, the distance between the starting address of consecutive blocks, equal to *csize*, will *not* be a power of two, leading to the starting address of subsequent blocks also being unaligned. Hence, all accesses will be unaligned, similar to the case of basic padding discussed in Section III.A.6). Similar performance patterns were also observed with other kernel configurations. The only anomaly we encountered was vector size of 8 for the R3W1 kernel where the unaligned cases achieved slightly higher performance than the aligned ones.

Fig. 14 shows the memory bandwidth efficiency (relative to theoretical peak) on both of our GPUs for the R1W1 kernel with varying halo sizes. The figure clearly shows that on GPUs, the performance is not only a lot more stable than our FPGA, but also the cache hierarchy on the GPUs absorbs the redundant memory accesses and increases the performance beyond what would be possible with pure reliance on the external memory. If we only consider the non-redundant accesses, an interesting pattern emerges. On the K20X GPU, performance variation is similar to that of Fig. 12 for small halo sizes. But as the halo size increases, the cache hit rate decreases and hence, performance does not increase relative to the increase in redundancy. On V100, however, the cache behaves more efficiently and the non-redundant performance closely follows the pattern of Fig. 12 where maximum efficiency is achieved if all accesses are at least 256-bit-aligned. It should be noted that since we use a smaller *bsize* (= work-group size) on K20X, the amount of redundancy on this GPU is higher than V100 for a fixed halo size.

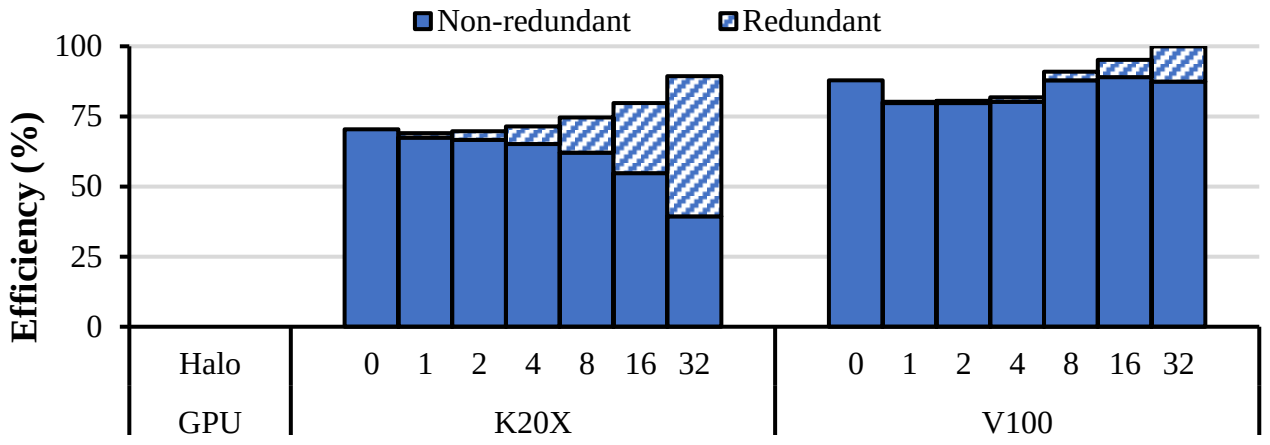

Fig. 14. Effect of overlapping on GPU performance for R1W1 kernel

*2) Effect of Padding:* In our previous work [4, 5, 6] we showed that padding can be used to improve memory access alignment when overlapped blocking is used. Fig. 15 shows the effect of different padding sizes on memory bandwidth efficiency with different halo sizes for the R1W1 kernel with a vector size of 16 and interleaving disabled.

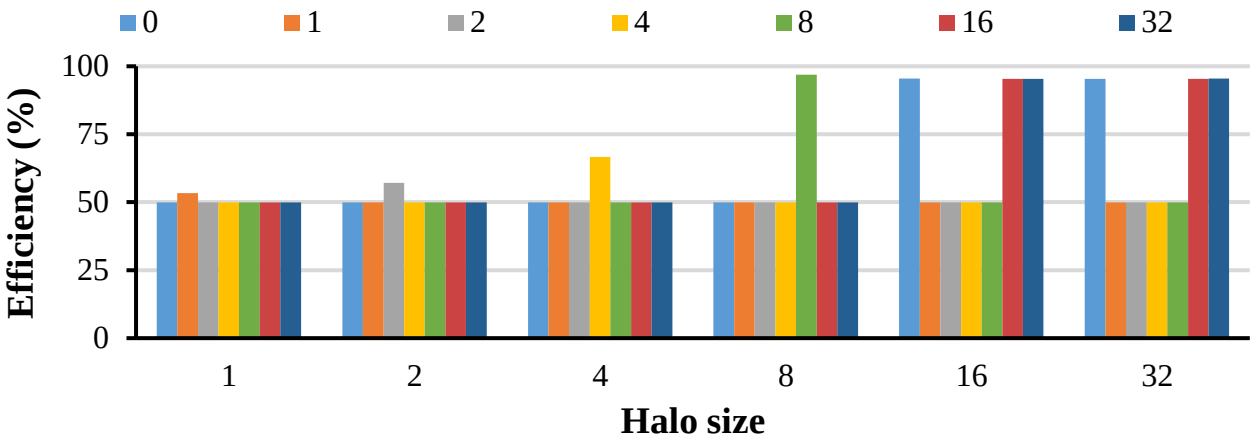

Fig. 15. Effect of padding on overlapped blocking with different halo sizes

A vector size of 16 requires a 512-bit alignment for maximum efficiency (Section III.A.6)). This requires the starting address of the first block to be 512-bit-aligned, and the distance between consecutive blocks (*csize*) be divisible by 512 bits so that all accesses are 512-bit-aligned. The first restriction requires a halo size divisible by 512 bits, and the second one requires divisibility by 256 bits. As Fig. 15 shows, when the halo size is divisible by the vector size, maximum performance can be achieved without padding. For a halo size of 8 (half the vector size), the second requirement for full alignment is already met, and the first requirement can also be met by padding the array with half the vector size, leading to maximum performance. For other halo sizes, it is impossible to meet the requirements for full alignment with padding; though, using a padding size equal to the halo size can still improve performance since it will move the starting point of some of the blocks to an aligned address. This pattern exactly follows what we previously reported with overlapped blocking for stencil computation on Intel FPGAs [4, 5, 6].

Fig. 16 shows the effect of padding for the R1W1 on our GPUs for varying halo sizes. For brevity, we limited the halo sizes to power of two up to 8, since higher values can achieve maximum memory bandwidth efficiency without padding (Fig. 14), and limited padding to two values: one is the same size as the halo size which achieves maximum memory bandwidth efficiency, and the other is 0, as a representative of every other padding size. Similar to the case of FPGAs, maximum memory bandwidth efficiency can be achieved on our GPUs with padding if memory accesses already have half the required alignment without padding (128 bits for GPUs) and improvements are also observed in other cases. The stark contrast between our GPUs and FPGA is that on Tesla K20X and V100, even without the help of padding, maximum performance loss due to memory access misalignment is 5% and 13% respectively, while on the FPGA, even with the help of padding, performance loss can reach as high as 45%. Moreover, on the GPUs, the required alignment size is fixed at 256 bits, while on the FPGA it can reach the more restrictive value of 512 bits depending on vector size.

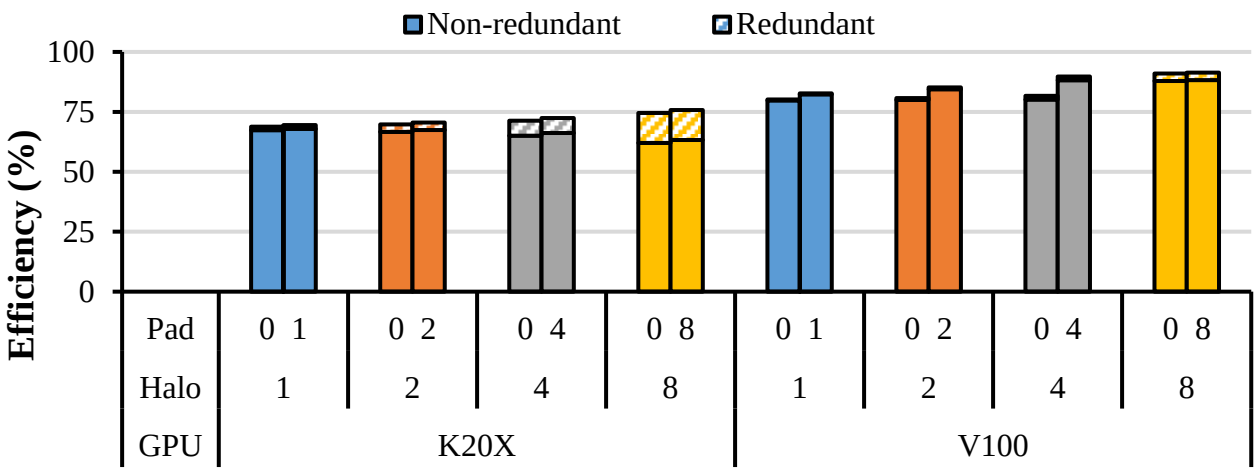

Fig. 16. Effect of padding for overlapped blocking on GPUs

*3) Effect of Cache:* Intel FPGA SDK for OpenCL Offline Compiler will generate a cache for burst-coalesced global memory accesses in a kernel "when the memory access pattern is data-dependent or appears to be repetitive" [2]. Testing with different versions of the compiler showed that this cache is generated for the Single Work-item variation of our kernels up to v17.0.2, but never for NDRange, while newer versions do not generate the cache at all despite the obvious overlapping of memory accesses. Using v16.1.2, we compared the performance of our kernels with the cache enabled or disabled and, despite the simple pattern of overlapping and the large Block RAM overhead of the cache, we observed no tangible performance difference which points to a critical disadvantage of the FPGA compared to the GPUs.

## *C. 1.5D and 2.5D Blocking*

Fig. 17 shows the FPGA bandwidth efficiency for 1.5D and 2.5D blocking using the R1W1 kernel with a vector size of 16 and interleaving disabled and varying halo (different clusters) and padding sizes (different bar colors). The halo size of zero in these figures represents bandwidth efficiency with non-overlapped blocking used in applications such as general matrix multiplication. For the 1.5D blocking case, we initially chose the *x* dimension to be the nearest multiple of *csize* to 16384; however, we observed unexplainable performance slow-down for vector sizes of 16 and 32 which only happened with specific multiples of *bsize* above 8192 (and their nearest multiple of *csize*), with no clear pattern. We suspect this could be yet another performance bug in the compiler/memory interface. In the end, we chose a size of 18432 for the *x* dimension to avoid this issue, and 16384 for the *y* dimension, resulting in a buffer size of ~1152 MiB. For the 2.5D blocking case, we fixed the size of the *z* dimension to 256, and chose both *x* and *y* to be equal to the nearest multiple of *csize* to 1024.

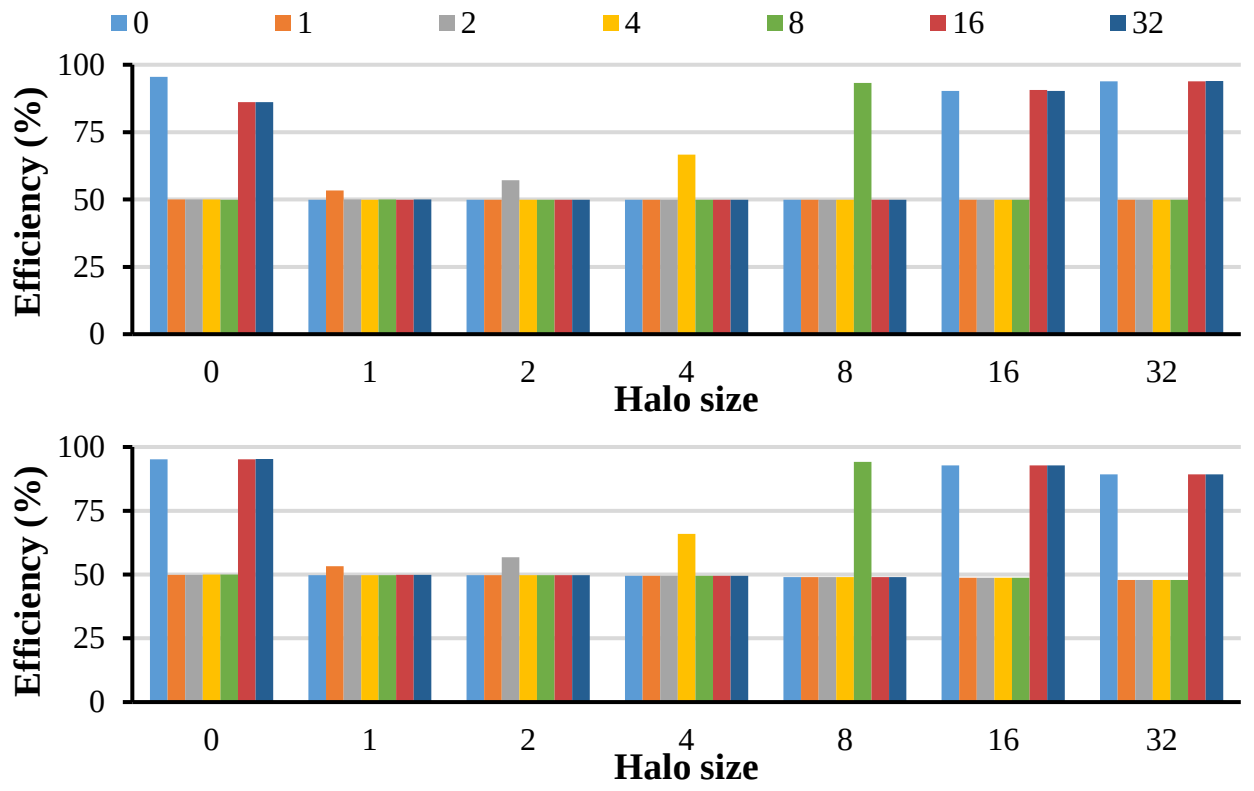

Fig. 17. Overlapped blocking efficiency; top: 1.5D, bottom: 2.5D

Both of these figures closely resemble the patterns seen in Fig. 13 and Fig. 15, showing that the memory access pattern has little effect on bandwidth efficiency as long as alignment characteristics remain unchanged. However, a few anomalies are observed here. For 1.5D blocking with a halo size of 0, padding sizes that are a multiple of 16 achieve 10% lower performance than no padding, which is counter-intuitive since they result in the same alignment. Though this anomaly is not a source of concern since padding is not required in this case. Efficiency is also ~5% lower than the achievable maximum value when halo size is 16 and enough padding is used to allow fully-aligned accesses, while this issue does not exist when halo size is 32. These anomalies are not observed with other vector sizes and their cause is unclear. For the 2.5D case, we observe that performance gradually decreases as the halo size gets larger which also happens with other vector sizes. Our benchmark avoids the out-of-bound rows and columns on the sides of the grid when overlapping is used (Fig. 1), and indexes falling in these areas are not counted for measuring memory bandwidth. However, since the compiler generates the pipeline in a way that the minimum latency of global memory accesses is absorbed without stalling the pipeline, these memory accesses, even though they are skipped, still incur that minimum latency and hence, they have the same effect on performance as a valid memory access, minus the possibility of stalling the pipeline. For 2.5D blocking, since the difference between *bsize* and *halo* is smaller, as the halo gets larger, the latency overhead of these uncounted accesses will increase relative to total run time, resulting in minor reduction in measured performance. The same effect can also be seen with 1D and 1.5D blocking and very large halo sizes. It is worth noting that in 1.5D and 2.5D blocking, memory access alignment also depends on the size of the *x* (for both) and *y* (for 2.5D) dimensions. For cases where these values are not a multiple of the vector size, row-wise and column-wise padding can be used to correct the alignment.

Our results here prove that the model accuracy values we reported in our previous work on stencil computation using overlapped blocking [4, 5, 6] were indeed the memory controller efficiency. The ~85% model accuracy reported in [4, 5, 6] for 2D stencils was the result of the compiler bug we reported in Section III.A.4), and converting the memory read/write kernels to NDRange allowed us to achieve ~100% of the predicted performance. Moreover, the 55-60% model accuracy for Diffusion 3D benchmarks with a vector size of 16 in the same works is close to the memory bandwidth efficiency values we reported in Fig. 17 for a halo and padding size of 4, which has the same alignment characteristics as the configuration of those benchmarks. The small difference between these values is due to the effect of non-linear performance scaling with operating frequencies above the memory controller frequency at large vector sizes (Section III.A.5)), and the aforementioned latency of out-of-bound memory accesses which are skipped. The performance bug caused by on-chip channels (Section III.A.4)) also applies to these cases, but for the configurations reported in [4, 5, 6], the performance loss caused by memory access misalignment hides the effect of this bug and using NDRange for the memory kernels does *not* improve the performance anymore. **One important point to note is that in many applications that rely on overlapped blocking, the halo size is *not* a configurable parameter. Hence, if the halo size is a value that does not allow maximum memory bandwidth efficiency with padding, there might be no straightforward way to improve memory performance except by modifying the memory controller itself to buffer and realign the memory accesses at run-time.**

As a final comparison, Fig. 18 shows the performance of the R1W1 kernel on both of our GPUs with different halo and padding sizes. The block size for the 1.5D blocking case is 1024, and for the 2.5D blocking case, even though slightly higher performance could be achieved using a block size of 16×16, we used 32×32 so that we could evaluate halo sizes up to 8 (*halo* must be smaller than $bsize/2$). Here, the variation of performance with respect to alignment is similar to Fig. 16, with the main different being lower performance improvement on both GPUs with caching, due to more complex access pattern and lower cache hit rate, especially with 2.5D blocking.

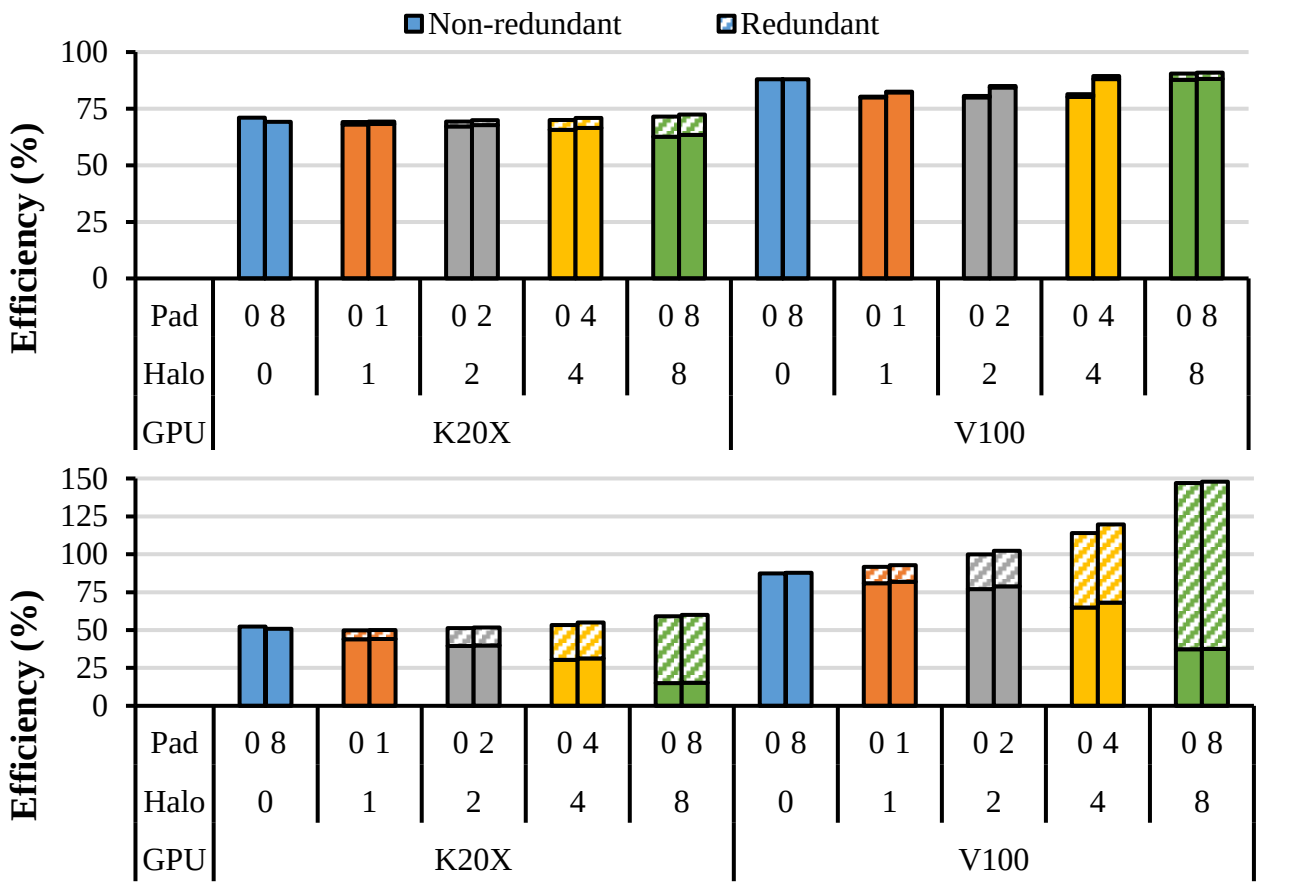

Fig. 18. Overlapped blocking efficiency on GPUs; top: 1.5D, bottom: 2.5D

## IV. Related Work

The most well-known memory benchmark in HPC is STREAM [12]. BabelStream [13] is a popular implementation of this benchmark with support for different programming languages and devices. However, it does not support FPGAs and only provides a small subset of the functionality of our benchmark suite. Our benchmark suite is specifically designed for FPGAs and covers a wider range of array configurations and memory access patterns with control over FPGA-specific parameters such as operating frequency, vector size, interleaving, etc. BabelStream covers more *computation* patterns than our suite, but the pattern of computation only affects *pipeline latency* on FPGAs, which has negligible effect on memory performance as long as the input size is large enough to saturate the pipeline. In [14], the authors present an implementation of the STREAM benchmark for FPGAs using OpenCL with support for continuous and strided memory patterns, and evaluate it with different buffer and vector sizes on Xilinx and Intel FPGAs. This work uses an early version of the Intel compiler, ignores the important effect of operating frequency variability, and does not explore the source of performance variations. We perform a more comprehensive analysis with different array, vector size, and alignment configurations, and discuss the performance trends in depth. Another port of the STREAM benchmark for FPGAs is also available at [15], and results for a Stratix 10 board with four DDR4 memory banks are reported. Their results confirm that also on Stratix 10, interleaving does *not* allow performance scaling to more memory banks than there are global buffers in the kernel. Moreover, they achieve a maximum bandwidth efficiency of ~85% with interleaving disabled (relative to number of banks used) which is ~10% lower than Arria 10.

## V. Conclusion.

In this work we presented a comprehensive analysis of the memory controller and memory bandwidth efficiency of Intel FPGAs using OpenCL. Our results showed that to achieve maximum memory performance, very strict requirements need to be met which only apply to a small portion of real-world applications and in typical cases, it might not be possible to achieve more than 70% of the peak memory bandwidth. Moreover, we showed that the memory controller is very sensitive to memory access alignment and up to 50% of the memory bandwidth might be lost in typical applications that rely on overlapped blocking due to unaligned accesses. Even though array padding can improve the performance in such cases, it is not always enough to achieve maximum performance, and major improvements in the memory controller itself will be required for Intel FPGAs to be competitive against GPUs in memory bandwidth efficiency.

## Acknowledgement

We would like to thank Andrei Vassiliev from Scientific Concepts International Corporation for pointing us to the right direction for overriding the operating frequency of OpenCL kernels. We would also like to thank Intel for donating FPGA software licenses through their university program. This work was performed under the auspices of the Real-world Big-Data Computation Open Innovation Laboratory, Japan.